\documentclass[pre,showpacs,twocolumn]{revtex4}
\usepackage[dvips]{graphics}
\usepackage{epsfig}

\newcommand{\g}{\gamma}

\newcommand{\lgl}{\langle}
\newcommand{\rgl}{\rangle}

\begin{document}

\title{Derivation of the percolation threshold for the network model of Barab\'asi and Albert}
\author{Wolfgang \surname{Pietsch}}
\email{wpietsch@gmx.de}
\affiliation{Department for Philosophy of Science, University of Augsburg, Universit\"atsstrasse 10, D-86135 Augsburg, Germany}

\begin{abstract}
The percolation threshold of the network model by Barab\'asi and Albert (BA-model) [Science 286, 509 (1999)] has thus far only been `guessed' based on simulations and comparison with other models. Due to the still uncertain influence of correlations, the reference to other models cannot be justified. In this paper, we explicitly derive the well-known values for the BA-model. To underline the importance of a null model like that of Barab\'asi and Albert, we close with two basic remarks. First, we establish a connection between the abundance of scale-free networks in nature and the fact, that power-law tails in the degree distribution result only from (at least asymptotically) linear preferential attachment: Only in the case of \emph{linear} preferential attachment does a minimum of topological knowledge about the network suffice for the attachment process. Second, we propose a very simple and realistic extension of the BA-model, that accounts for clustering. We discuss the influence of clustering on the percolation properties.\end{abstract}


\maketitle
\pagestyle{headings}

\section{Introduction}

Scale-free networks, i.e., networks with essentially power-law degree distributions, have recently been widely studied (see \cite{baretalb} and \cite{dorogov} for reviews). Such degree distributions have been found in many different contexts, for example in several technological webs like the Internet \cite{int}, the WWW \cite{WWW}, or electrical power grids, and also in social networks, like the network of sexual contacts \cite{sex} or of phone calls \cite{phonecall}.

The standard model reproducing scale free degree distributions was introduced by Barab\'asi and Albert (BA-model) \cite{BA-model}. It is based on a growth algorithm with preferential attachment. Starting from an arbitrary set of initial nodes, at each time step a new node is added to the network. This node brings with it $m$ proper links which are connected to $m$ nodes already present. The latter are chosen according to the preferential
attachment prescription: The probability that a new link connects to a certain node is proportional to the degree (number of links) of that node.
The resulting degree distribution of such networks tends to \cite{degdis}
\begin{equation}
P(k)=\frac{2m(m+1)}{k(k+1)(k+2)} \propto k^{-3}.  \label{degdis}
\end{equation}

A second older model which is also widely studied in the context of scale-free networks is the configuration model (C-model). This model is usually attributed to Bollob\'as \cite{bollobas} and was first treated in a context related to percolation by Molloy and Reed \cite{mollreed}. This is to some extent the `most random' network possessing a given degree distribution $P(k)$ and a given number $N$ of nodes. The building prescription starts with sets of $NP(k)$ nodes with $k$ stubs each. The stubs are then connected randomly to each other; two connected stubs form a link. Double bonds and autoconnections can be neglected in the limit of large networks $N \rightarrow \infty$.

\section{Percolation condition}

\subsection{Theory}
\label{sec:pcm1}
Let us recall the condition for a network at the percolation threshold \cite{cohetal, cohetal2}: A node $i$, linked to a node $j$ in the spanning cluster, is connected to exactly one other node on average. This results in an average degree $\lgl k \rgl=2$ of the spanning cluster. These properties are obviously properties of trees. More precisely, if the following conditions are satisfied, a network is at the percolation transition:
\begin{enumerate}
\item There exists a giant cluster, which is a tree (i.e., no loops).
\item The distance between two randomly chosen nodes is almost always infinite (in the thermodynamic limit $N \rightarrow \infty$). That is, the fraction of pairs, for which it is not infinite, is zero.
\end{enumerate}
We prove this in the following manner: On a tree [condition (1)] there is always exactly one path between two nodes. Now, we randomly delete from the tree a fraction $p$ of all links or nodes, corresponding to edge or site percolation respectively. Due to condition (2), the probability that two arbitrarily chosen nodes still belong to one cluster is zero independent of $p>0$: It is $(1-p)^n$, where $n$ is the distance (number of links) between the two nodes, and since the distance diverges for almost all pairs of nodes, the probability, that the nodes are connected is zero. In consequence, there does not exist any cluster that consists of a non-zero fraction of nodes, i.e., no giant cluster exists.

\subsection{Application to the BA-tree}
The BA-network with $m=1$ (BA-tree) has two peculiar qualities: first it is exactly a tree, second it is fully connected, i.e., it consists of one single cluster (iff the starting network is fully connected -- otherwise there are as many giant components as there were components in the starting network at $t=0$). Using the conditions introduced above, we now examine, if $p_c=0$ is the percolation threshold.

Obviously, condition (1) is fulfilled by the construction algorithm. Condition (2) also holds. In \cite{ultrasmall} Cohen and Havlin calculate a lower limit for the diameter (mean distance between all nodes) of scale-free networks. For a given degree distribution they build a tree by starting with the node of the highest degree as root, and then subsequently adding as offspring the nodes with the next-highest degrees. When a shell is full, nodes are added to the next shell etc.\ (CH-model) (To the same shell belong all nodes with a fixed distance from the root.) Cohen and Havlin find that the number of shells diverges in the case of the distribution of the BA-model. This of course does not immediately entail the divergence of the distance between almost all pairs of nodes.

To prove the latter for the CH-model, it suffices to consider pairs including the node with the highest degree, because it has the maximum number possible of neighbors in all shells. Now, from the construction algorithm of the CH-model as described above we know that on every shell there are at least as many nodes than on the previous (excluding the last shell, because nodes on the penultimate shell can have degree one). This together with the fact that the number of shells diverges, proves that the distance between the node with the highest degree and almost all other nodes diverges in the CH-model. Now, we consider all networks that are trees and that have the same degree distribution as the BA-model with $m=1$. When in any such network we arbitrarily choose one node as root and count the number of nodes in the shells around it, we find, that there are always less nodes in each shell than in the corresponding shell of the CH-model, where we take the node with the highest degree as root (except again the last shell). It follows that condition (2) must be fulfilled for the BA-model. 

We have shown, that the percolation threshold for the BA-model with $m=1$ is $p_c=0$. Of course, the percolation threshold for a C-network with the same degree distribution is $p_c=1$ due to the diverging second moment\cite{cohetal, cohetal2}. So, for the BA degree distribution with $m=1$, $p_c$ depends strongly on correlations.

\subsection{Simulations}
\begin{figure}
\epsfig{file=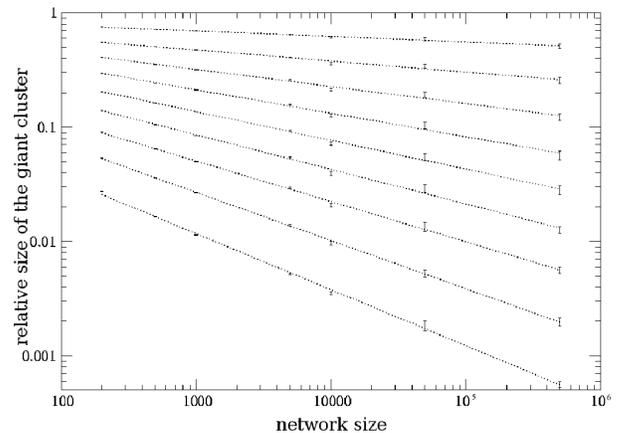, width=8.6cm}
\caption{\label{fig:simm1} Link removal in BA-networks with $m=1$. The scaling of the percolation process with increasing network size is examined. The nine curves correspond to $p=0.1$--$p=0.9$ in steps of $0.1$ from top to bottom. The values for the different network sizes were averaged over 9999 runs for 200, 500, and 1000 nodes, over 500 runs for 5000 nodes, over 100 runs for 10,000 nodes and 20 runs for both 50,000 and 500,000 nodes. The graph suggests that for infinite network size, there will be no giant cluster independent of $p>0$.}
\end{figure}
The simulations in Fig.\ \ref{fig:simm1} confirm the theoretical discussion from above. The graph documents the percolation process for different network sizes from 200 to 500,000 nodes. We plotted for different $p$ the relative size $s$ of the giant component depending on the network size. It is shown that for a fixed $p>0$ the relation between the relative size of the giant component $s$ and the network size can be described by a simple power-law with a negative exponent. Thus, for all $p>0$, the relative size of the giant component $s$ approaches zero for $N \rightarrow \infty$.

\section{The role of the tree structure}
\label{sec:tree}
We will now try to determine the origin of the percolation threshold $p_c=0$ for the BA-tree. As already mentioned, this threshold contradicts the common notion (proven for the C-model), that for power-law exponents $\g \leq 3$ the diverging second moment of the degree distribution yields a percolation threshold $p_c=1$. Does the unusual percolation threshold $p_c=1$ stem from the peculiarities of the BA-tree? In the following we examine the importance of the tree topology. For this purpose we determine the change of the percolation threshold when we randomly add a few links to the BA-tree. Therefore, we calculate the cluster-size distribution for the BA-tree depending on $p$.

\subsection{The cluster-size distribution for the BA-tree}
\label{cl-size}
In the BA-model each link has a direction determined by the preferential attachment rule: Each node has exactly $m$ incoming or proper links that are attached to it at the moment when the node enters the network. All other links will be called outgoing links. In the following we consider $m=1$, the connected tree. We remove a certain fraction $p$ of all links from the network. Then the network breaks into as many clusters as links are removed (plus one).
\begin{figure}[tb]
\begin{center}
\epsfig{file=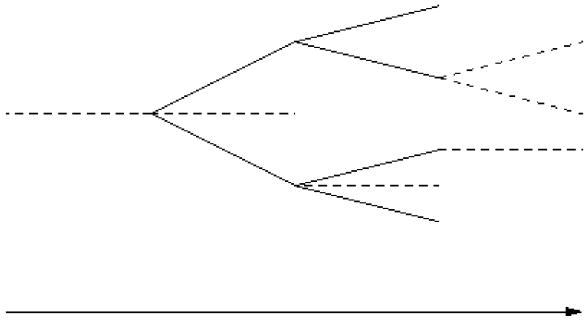, width=8.6cm}
\caption{An exemplary cluster. Each node has an incoming link in the direction of the arrow. All other links are counted as outgoing. The dotted lines are the removed links, the solid ones are the links drawn. On the far left we see the only removed incoming link for the whole cluster.}
\label{fig:exclus}
\end{center}
\end{figure}
For every cluster one incoming link was deleted and a certain number of outgoing links. In Fig.\ \ref{fig:exclus} an exemplary cluster is shown. There, links enter nodes in the direction of the arrow.

We are interested in the cluster-size distribution in the limit of large cluster sizes. First, we count the outgoing links of a cluster. $L_s$ drawn outgoing links will belong to this cluster. In the whole network, only $(1-p)$ of all links are drawn. Then, $L_u=x \cdot L_s$ deleted outgoing links belong to the cluster. In the limit of clusters with many nodes, we have $x=p/(1-p)$.

We now examine the development of a percolated network, i.e., already at the entry of a node the decision is made, if its proper link is drawn. We can formulate a dynamical equation for the evolution of the cluster sizes with a large number of nodes $k$:
\begin{eqnarray}
\frac{d}{dt}C(k)&=&(1-p)\left \{ \frac{C(k-1)\left [ (k-2)(2+\frac{p}{1-p}) \right ]}{2N} \right . \nonumber \\
&&-\left . \frac{C(k)\left [ (k-1)(2+\frac{p}{1-p})\right ]}{2N} \right \}.
\label{Equation2}
\end{eqnarray}
$C(k)$ is the number of clusters with $k$ nodes in the network. At every time-step a new node is added, whose link is drawn with probability $(1-p)$, which is the first factor on the right-hand-side of the equation. If the link is deleted, automatically a cluster of size $1$ is added to the network. We neglect this term since we are interested only in the limit of large $k$. The first addend on the right-hand-side accounts for the new node being linked to a cluster with size $k-1$. The number of drawn stubs is $2(k-2)$, and $(k-2)p/(1-p)$ is the number of deleted (outgoing!) stubs. (A stub is half a link; we neglect the single stub of the one deleted incoming link.) The sum of both yields the total number of stubs in the cluster that determines the linear preferential attachment of a new node to the cluster. $2N$ is approximately the total number of stubs in the network with $N$ nodes. The second addend accounts for the new node being linked to a cluster of size $k$. 

\begin{figure}[tbp]
\begin{center}
\epsfig{file=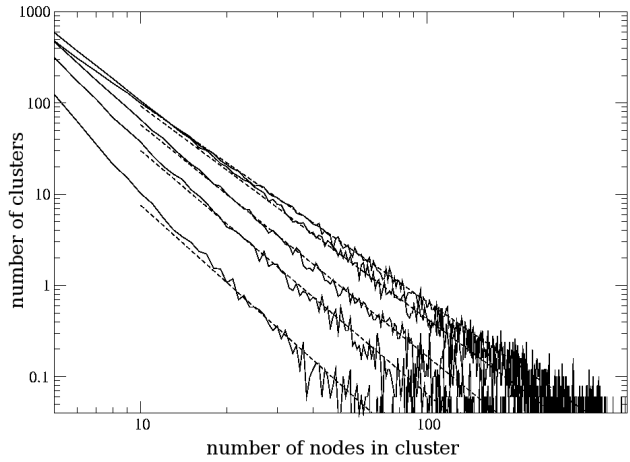, width=8.6cm}
\caption{\label{fig:ckres} Distribution of cluster-sizes in BA-trees with 100,000 nodes each. The five curves correspond to $p=0.3$, $0.5$, $0.7$, $0.8$, and $0.9$ (from top to bottom, each averaged over 50 runs). The dashed lines show the corresponding theoretical results for the exponent of the power-law tail, resulting for a double-logarithmic plot in a line with slope $-1-1/(1-0.5p)$. The offset for the theoretical curves is adjusted manually.}
\end{center}
\end{figure}

We now assume that $C(k)=c(k)t = c(k)N$ with $c(k)$ constant for large $N$, where $N$ is the number of nodes in the network. (This follows immediately from the statistical nature of the networks.) We have
\begin{eqnarray}
c(k)&=&(1-p) \left \{ c(k-1) \left [ (k-2) \left (1+\frac{1}{2} \frac{p}{1-p} \right ) \right ]  \right . \nonumber \\
&&- c(k) \left . \left [ (k-1) \left (1+\frac{1}{2}\frac{p}{1-p} \right) \right ] \right \}.
\end{eqnarray}
We can solve this equation for $c(k)/c(k-1)$ with $k \gg 1$:
\begin{equation}
\frac{c(k)}{c(k-1)}=\frac{k-2}{k-1+\frac{1}{1-\frac{1}{2}p}}.
\end{equation}
In the limit of large $k$ follows
\begin{equation}
c(k)\propto k^{-\left (1+\frac{1}{1-\frac{1}{2}p} \right )}.
\label{eq:ckres}
\end{equation}
We see that for $0 < p < 1$, the exponent lies between 2 and 3. We checked Eq.\ (\ref{eq:ckres}) in Fig.\ \ref{fig:ckres}, where we plotted the
cluster-size distributions for BA-trees with 100,000 nodes.
For five values between 0.3 and 0.9 we found a good correspondence between the simulated curves and the theoretical prediction.

\subsection{Randomly adding links to the BA-tree}
\label{random}
The calculation of the cluster-size-distribution is not easily generalized to BA-models with $m>1$. We will now destroy the tree topology by adding links between randomly and uniformly chosen nodes. We will call these links R-links. Note, that this does not change the exponent of the power-law tail of the degree distribution. Does it change the percolation threshold? The new links connect the `old’ clusters, which are distributed according to $C(k)$. The number of R-links added to a cluster is for large clusters proportional to the cluster size $k$, i.e., the number of nodes in the cluster. 

Let $r$ be the average number of links added to every node. Now, it can be shown that as long as $r>0$ a spanning cluster exists that comprises a finite fraction of all nodes in the network. The argument is analogous to the reasoning why there is a spanning cluster in a C-network with diverging second moment of the degree distribution. When following one of the newly added R-links, we find the following probability distribution for encountering a cluster of size $k$: $k C(k)/[\sum_k k C(k)]$. For large enough $k$, the total number of R-links connecting to such a cluster of size $k$ will approximately be $rk$. Now, when we follow an R-link to a cluster, we can calculate the average number of outgoing R-links from this cluster:
\begin{equation}
r\frac{\sum\limits_k k (k-1) C(k)}{\sum\limits_k k C(k)}.
\label{eq:superclus}
\end{equation}
If this expression diverges a finite fraction $>0$ of all clusters will be connected after the addition of the R-links (for the same reason that in the C-model with diverging second moment of the degree distribution a spanning cluster develops). To this `super-cluster' clusters with a large number of nodes will belong with a higher probability than small clusters and it follows directly that the super-cluster also comprises a finite fraction $>0$ of the nodes in the network, i.e., the super-cluster is also a spanning cluster.

We calculated in Sec.\ \ref{cl-size} that for all $p$ between 0 and 1 the cluster-size distribution $c(k)$ of the BA-tree has a power-law tail with exponent between 2 and 3. Since for these $c(k)$ the expression (\ref{eq:superclus}) diverges, there always exists a giant component (the super-cluster mentioned above) in the BA-tree with additional random links  independent of $r>0$. That leads to a percolation threshold $p_c=1$ for the model with additional links opposed to $p_c=0$ for the BA-tree. In this sense the BA-tree has a critical topology.

\subsection{Mapping the $m=1$ BA-model onto a $m=2$ BA-model}
\label{sec:3c}
In this section we prove that the BA-model with $m>1$ has a percolation threshold of $p_c=1$. There exists an easy way of mapping the BA-model with $m=1$ onto the BA-model with $m=2$. The prescription is the following (cp. Fig.\ \ref{fig:mapm1m2}):
\begin{figure}[tb]
\begin{center}
\epsfig{file=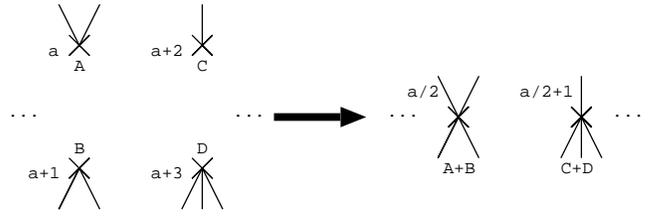, width=8.6cm}
\caption{Mapping $m=1$ (left side) on $m=2$ (right side) networks. Pairs of nodes with difference in age of one are combined into one single node. $a$ etc.\ denote here the age of the nodes.}
\label{fig:mapm1m2}
\end{center}
\end{figure}
\begin{enumerate}
\item Partition the whole network into pairs of nodes with a difference in age of one (the age of a node is a natural number representing the moment when the node is added to the network, nodes are numbered consecutively).
\item Replace each pair of nodes by a single node.
\end{enumerate}

With the same effect we can add a `non-removable' link between the two nodes of each pair (`non-removable' links are not affected by percolation). With the addition of the `non-removable' links we add to every node of the BA-tree one additional link -- independent of $0 \leq p \leq 1$. When translating the number of nodes in a cluster of the BA-tree with additional links to the number of nodes in a cluster of the BA-model with $m=2$, we have to divide the number of nodes by $2$, because the entity of two nodes and one `non-removable' link corresponds to one single node in a BA-network with $m=2$.

To apply the considerations of the last section \ref{random} in order to prove that $p_c=1$ for a BA-model with $m=2$, we still have to show that the `non-removable' links between the nodes are randomly distributed. Actually it suffices to consider only large clusters of size $k>k_0$ with an arbitrary but finite $k_0$. (Those large clusters comprise a non-zero fraction of all nodes in the network.) Because: if there is a giant component in the subnetwork of those large clusters, then there is also a giant component in the whole network, since the large clusters already constitute a finite part of the network.

Due to the statistical nature of our networks, we can say that in all large clusters we find approximately the same age distribution. For this reason we can say that the large clusters are randomly linked to each other and that the probability for connecting to a cluster j with one `non-removable' link (originating in cluster i) is proportional to the number of nodes in cluster j. Thus, the network of the large clusters corresponds to the network-model treated in the last section \ref{random}. Since in the subnetwork of large clusters, the cluster-sizes are distributed according to a power-law with an exponent $\leq 3$, there exists a giant component in the whole network (as we have proven in the previous section). 

Who is not convinced by this statistical argument, consider a percolated BA-network of size $N$ with $m=1$ and a certain $p$. In such a network we observe a certain cluster-size distribution according to Eq.\ (\ref{eq:ckres}). We now let the network develop to size $1.5 N$ with $m=2$, i.e., from now on every new node shall have two proper links. If both proper links are drawn [probability $(1-p)^2$], then the new node will link two clusters in the original network of size $N$ with a probability larger than $1/4$ (lower limit for the probability that both links connect to the original network). Since according to preferential attachment the new links will connect to existing clusters only dependent of the cluster-size, these new nodes will serve as bridges between clusters exactly in the way required in Section \ref{random}. Thus, independent of $p<1$, BA-networks with $m=2$ will have a giant cluster. q.e.d.  

This consideration can easily be generalized to BA-networks with other $m>1$. We can even treat networks with fractional $\lgl m \rgl$, where $\lgl m \rgl$ is the average number of additional links added per step (with an upper boundary $m_0$ for the number of links added in each step). For $\lgl m \rgl >1$ the percolation threshold is always $p_c=1$.

\section{A note on scale-free nature and linear preferential attachment}
\label{sec:sim}

In the following we will briefly address a few essential aspects of the BA-model. These underline the importance of this model as a null model. First, we will establish a connection between the abundance of scale-free networks in nature and the fact, that scale-free degree distributions emerge only for asymptotically \emph{linear} preferential attachment. For linear preferential attachment only minimal topological knowledge about the network is required for attaching a new node to the network. 

It is well known that non-linear preferential attachment results in degree distributions, that are not scale-free \cite{Kra}. In the sublinear case the degree distribution is a stretched exponential. In the asymptotically linear case, the degree distribution follows a power-law asymptotic behavior for large degrees $k$. In the super-linear case, a `winner takes all' phenomenon arises, i.e., a single dominant gel node emerges.

We now show, why linear preferential attachment is indispensable for an economy in the information required for the attachment of a new node. Only for linear preferential attachment can the new node connect in a hierarchical process, subsequently increasing the resolution of only the portion of the network that the node will attach to.
 
Consider somebody who wants to publish a new webpage about a certain topic, e.g. India. To which already existing pages does he link his page? There are millions of webpages about India, that can be categorized into several topics. Our India-fanatic chooses a few of those topics (first level): For example, he wants to make a link to a travel agency that organizes tours through India. Then, he chooses one of the agencies according to their whole Internet representation, combining several webpages (second level). Following his choice, he will look in detail at the web-pages of the agency, and will then decide to which page exactly he makes his link (third level). We will find in the following that this hierarchical process of decision-making is only possible with \emph{linear} preferential attachment.
 
We introduce the component transformation that lets us combine arbitrary groups of nodes to hyper-nodes. The component transformation allows to look at the network at different resolutions. At lower resolutions, the amount of information necessary for the characterization of the network is smaller. The component transformation allows that a new node attaching to the network can do so in a hierarchical process, consecutively increasing the resolution of only the part of the network, it attaches to. We find that this hierarchical connection procedure, which follows the same preferential attachment rule at every resolution, is only possible for networks with linear preferential attachment.

We define the component transformation as shown in Fig.\ \ref{fig2}:

\begin{figure}
\includegraphics[width=8.6cm]{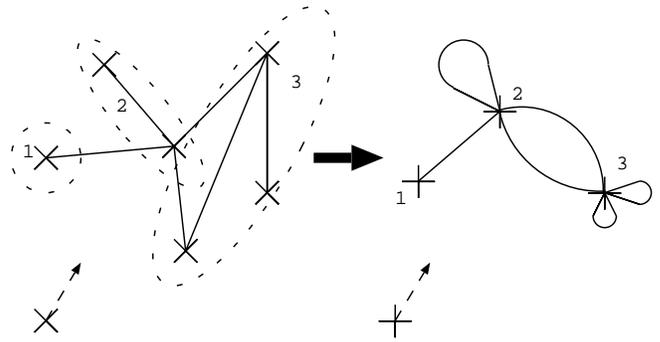}
\caption{The component transformation. Arbitrary nodes are grouped together to form a single node. All outgoing edges are preserved in the process. Edges connecting nodes within one group are converted into autoconnections.}
\label{fig2}
\end{figure}
\begin{enumerate}
\item Nodes are arbitrarily grouped into neighborhoods, so that each node can be assigned to exactly one neighborhood.
\item Each neighborhood is replaced by a hyper-node.
\item Links between nodes of different neighborhoods are converted into links between the corresponding hyper-nodes.
\item Each link between nodes of the same neighborhood is converted into an autoconnection of the corresponding hyper-node.
\end{enumerate}
Using the component transformation we can consider each network at many different levels/resolutions. After a component transformation the network can again be subjected to another component transformation. In each transformation, information is lost. However, for a new node $i$ this information is irrelevant for the decision, to which \emph{hyper-node} $j$ it will connect according to preferential attachment. 

Then, to attach node $i$ to the network it suffices to look at the fine structure only of the hyper-node $j$, while the fine structure of all other hyper-nodes is irrelevant for the attachment process. Finally, the fine structure of the hyper-node can again have a superfine structure etc.
 
We now prove, that this hierarchical process is only possible for \emph{linear} preferential attachment
\begin{equation}
\Pi_i=\frac{k_i}{\sum\limits_{n} k_n}.
\end{equation}
Here, the probability $\Pi_i$, that a new node attaches to the old node i, is proportional to the degree of that node $k_i$. The component transformation is $\sum_{I_i} i \mapsto I$, i.e., the nodes i are combined to groups $I_i$, which are replaced by the hyper-node $I$. We also have $\sum_{I_i} k_i\mapsto k_I$, i.e., the degree of the hyper-node $k_I$ equals the sum of degrees of all nodes in $I_i$. Now, with linear preferential attachment the probability, that a new node attaches to $I$, is
\begin{equation}
\Pi_I=\frac{k_I}{\sum\limits_{I} k_I}=\frac{\sum\limits_{I_i} k_i}{\sum\limits_{I}\sum\limits_{I_i} k_i}=\sum\limits_{I_i}\Pi_i.
\end{equation}
So, the probability that a node attaches to a group of nodes $I_i$ is equal to the sum of the probabilities that it attaches to one of the nodes in the group. We stress, that this does not work anymore with non-linear preferential attachment:
\begin{equation}
\Pi_I=\frac{k_I^{\alpha}}{\sum\limits_{I} k_I^{\alpha}}\neq\frac{\sum\limits_{I_i} k_i^{\alpha}}{\sum\limits_{I}\sum\limits_{I_i} k_i^{\alpha}}=\sum\limits_{I_i}\Pi_i.
\end{equation}
So, we have shown, that linear preferential attachment assures that a new node does not have to know the fine structure of the whole network in order to `decide', which part it will attach to. The new node does not need to know the topology of the whole network, and it does not need to know at which resolution it is looking at the network. This scale-independent quality of the attachment procedure only works for linear preferential attachment and is lost for non-linear preferential attachment. We have shown that requiring a minimization of the amount of information available to the new node results in \emph{linear} preferential attachment, and thus in a scale-free degree distribution. These considerations can easily be generalized to linear preferential attachment with initial attractiveness $A_i\neq 0$ of node i. The hyper-node $I$ then simply has the attractiveness $A_I=\sum_{I_i} A_i$. An essential quality for the preferential attachment seems to be, that the value/worth attributed to a larger portion of a network is equivalent to the sum of the values of its parts.

\section{Clustering and the BA-model}
The most problematic aspect of the BA-model is the lack of clustering, which stands in harsh contrast to observations in real networks. We will in the following propose a very simple extension of the BA-model, that allows to implement a wide range of clustering, while it exactly preserves the degree distribution of the BA-model.

We assume that every new node added to the network brings with it $m=2$ proper links. These proper links connect the new node with nodes in the network according to different criteria. For example, in a friendship-network, every individual would have the right to make two friends. The first friend he chooses from people who do the same job as he. The second friend he chooses from people who have the same favorite hobby. Both times he preferably befriends those people that already have a lot of friends (i.e., preferential attachment).

The new feature compared with the BA-model is, that at its introduction we assign to each node $i$ two parameters, a job-parameter $0 \leq p_{j,i} \leq 1$ and a hobby-parameter $0 \leq p_{h,i} \leq 1$. Each new node has a job-link and a hobby-link. Now, according to preferential attachment we first determine the degree $k_x$ of node $x$, that the job-link will attach to. Then we search for that node $x$ with degree $k_x$ that has the $p_{j,x}$ closest to $p_{j,i}$, corresponding to the best matching of common interests. The same procedure determines node $y$, that the hobby-link of $i$ attaches to. The parameters $p_j$ and $p_h$ can of course be identified with two-dimensional coordinates in a geography. A generalization to more than two parameters is straight-forward. 

Note, that by definition of the evolution-algorithm the degree distribution develops in the same way as for BA-networks, independent of the clustering effect. This can be seen on a step by step basis. Every time a new node and its two proper links are added, the existing degree distribution together with the preferential attachment procedure exclusively determines the further evolution of the degree distribution. The topology of the network plays no role. Thus, the degree distribution of our model including clustering will not be different from that of a simple BA-model.

Qualitatively, the clustering depends on the correlations between the parameters $p_{j,i}$ and $p_{h,i}$. As a rule, we choose $p_{j,i}$ uniformly at random. $p_{h,i}$ is chosen depending on the value $p_{j,i}$. There are two limiting cases:
\begin{enumerate}
\item The choice of $p_{h,i}$ is independent from the choice of $p_{j,i}$. Then our model corresponds exactly to the BA-model with $m=2$ and exhibits a very small clustering coefficient, that vanishes as $N \rightarrow \infty$. (The clustering coefficient for a single node $i$ is commonly defined as the number of direct neighbors of $i$, that are linked with each other, divided by the number of possible pairs of direct neighbors of $i$. The clustering coefficient of the whole network is the average of all clustering coefficients for the single nodes \cite{Wastro}.)
\item $p_{h,i}=p_{j,i}=p_i$. Then the clustering is maximal.
\end{enumerate}
For the second case, when the degrees $k_x$ and $k_y$ of nodes x and y are equal, a double bond would be formed. If as an additional rule we prohibit double bonds, the second link will be connected to a node $y$ with the parameter closest to $p_i$ but unequal node $x$. Now, the clustering is maximal, because the probability that nodes $x$ and $y$ are neighbors is maximal. In the case that $x$ and $y$ are neighbors a new triangle is formed in the network. Note, that in this case the clustering can be fairly independent of the network size.

Let us finally address the percolation properties of these clustered networks. It is clear, that for strong correlations between $p_h$ and $p_j$ as in the second case the argument presented in Section \ref{sec:3c} is not valid anymore. The reason for this is, that the additional undeletable links tend to link nodes in the same cluster. Mapped on a BA-model with $m=1$, job- and hobby-nodes would be added alternately to the network. Subsequent nodes would pairwise have the same parameter $p_i$, and would consequently be linked to the same neighborhood with a high probability. So, when we reconstruct the $m=2$ model, in many cases nodes would be combined, that already belong to the same neighborhood. In case of strong correlation between the hobby and job parameters, neighborhoods form with similar job- and hobby-parameters. However, as pointed out in Section \ref{random}, the peculiar percolation behavior $p_c=1$ will result, if just a few small-world links are added: For example if an arbitrarily small but finite fraction of hobby-parameters are chosen totally independent of the job-parameter, the percolation threshold $p_c=1$ would be recovered.

\section{Conclusion}

We have proven, that the percolation threshold of the Barab\'asi-Albert model is in fact $p_c=1$, except in the case $m=1$, when only one link is added for each new node. To our knowledge this was shown here for the first time in the physics literature. As pointed out by the anonymous referee, in \cite{bollrio} Bollob\'as and Riordan also briefly address, how from their derivation of the diameter of BA-networks follows the value for the percolation threshold of BA-networks with $m \geq 2$. Apart from that, thus far the value of the percolation threshold was based only on simulations and comparison with the configuration model. It is worth noting that the result achieved here is perhaps not as evident as it may seem. Due to the uncertain role of correlations, the BA-model and the configuration model are not equivalent in the percolation properties as can be seen from the case of the BA-model with $m=1$. We closed with remarks underlining the importance of the BA-model as a null model and how clustering can be naturally accounted for.
\\

\section*{Acknowledgments} 

The author thanks Igor M.\ Sokolov for fruitful discussions and gratefully acknowledges helpful comments by the anonymous referee.

\end{document}